\documentclass{article}

\usepackage{arxiv}

\usepackage[utf8]{inputenc} 
\usepackage[T1]{fontenc}    
\usepackage{hyperref}       
\usepackage{url}            
\usepackage{booktabs}       
\usepackage{amsfonts}       
\usepackage{nicefrac}       
\usepackage{microtype}      
\usepackage{lipsum}
\usepackage{color}
\usepackage{algorithm}
\usepackage{algorithmic}
\usepackage{amsmath,amsfonts,amssymb}
\usepackage{subfigure}
\usepackage{lineno}
\usepackage{graphicx}
\usepackage{bm}
\graphicspath{{figures/}}
\usepackage{subfigure}

\title{A cyclic perspective on transient gust encounters through the lens of persistent homology}

\author{Luke Smith\thanks{Corresponding author: lsmith1@ucla.edu}   ,
\hspace{0.1em}
Kai Fukami,
\hspace{0.05em}
Kunihiko Taira
\\
Department of Mechanical and Aerospace Engineering, \\
University of California, Los Angeles, CA 90095, USA \\
\\
{\bf Girguis Sedky\thanks{Current affiliation: Department of Mechanical and Aerospace Engineering, Princeton University, 08544}   ,
\hspace{0.01em}
Anya Jones
}
\\
Department of Aerospace Engineering, \\
University of Maryland, College Park, MD 20740, USA \\
}

\begin{document}
\maketitle

\begin{abstract}
Large amplitude gust encounters exhibit a range of separated flow phenomena, making them difficult to characterize using the traditional tools of aerodynamics. In this work, we propose a dynamical systems approach to gust encounters, viewing the flow as a cycle (or a closed trajectory) in state space. We posit that the topology of this cycle, or its shape and structure, provides a compact description of the flow, and can be used to identify coordinates in which the dynamics evolve in a simple, intuitive way. To demonstrate this idea, we consider flowfield measurements of a transverse gust encounter. For each case in the dataset, we characterize the full-state dynamics of the flow using persistent homology, a tool that identifies holes in point cloud data, and transform the dynamics to a reduced-order space using a nonlinear autoencoder. Critically, we constrain the autoencoder such that it preserves topologically relevant features of the original dynamics, or those features identified by persistent homology. Using this approach, we are able to transform six separate gust encounters to a three-dimensional latent space, in which each gust encounter reduces to a simple circle, and from which the original flow can be reconstructed. This result shows that topology can guide the creation of low-dimensional state representations for strong transverse gust encounters, a crucial step toward the modeling and control of airfoil-gust interactions.
\end{abstract}

\section{Introduction}
\label{sec:introduction}

\begin{figure}
\centerline{\includegraphics[width=\textwidth]{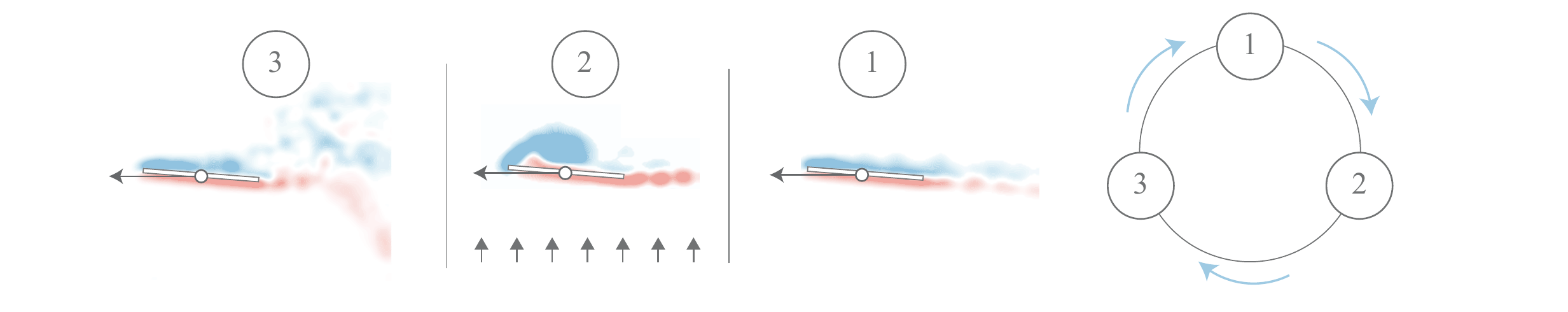}}
\caption{The broad stages of vortex shedding associated with a transverse gust encounter.}
\label{fig1}
\end{figure}

\noindent In an increasing number of flight applications, aerodynamic bodies are subject to large, unsteady disturbances, often resulting in separated, vortex-dominated flows~\cite{jonesCetinerSmith2021}. These flows are difficult to model in a low-order manner, but conceptually, vortex shedding events can often be described in intuitive terms. Consider the problem of a wing encountering a discrete gust, illustrated in figure~\ref{fig1}. This flow evolves in roughly three stages: (1) the wing begins in a base state of cruise, likely characterized by attached flow; (2) the wing encounters the gust, which triggers vortex shedding from the leading edge; (3) the wing exits the gust and returns toward its base state. This flow is not periodic and the underlying physics are inherently nonlinear, but in broad terms, these stages of vortex shedding belong to a single cycle, from the base state to a disturbed state and back to the base state. 

The current work poses the following question: is there a way to describe large-amplitude disturbances that appeals to our intuitive notion of cycles and loops? Cyclic events in aerodynamic flows have historically been described using Fourier analysis, but for discrete transient maneuvers, in which the dynamics of the flow are unlikely to conform to a time-periodic basis, it can be difficult to characterize the flow without resorting to a large number of modes. As an alternative, we propose a data-driven framework that nonlinearly characterizes disturbed flows based on their topology, or their shape and structure, in a high-dimensional state space. Our central idea is that large-scale vortex shedding events, while complex and nonlinear in physical space, exhibit a fairly simple topology in state space, which we can leverage to identify low-order, interpretable representations of the flow. 

To demonstrate this approach, we consider a set of experimental flowfield measurements in which a flat-plate wing translates horizontally into a transverse gust. For each gust encounter, we analyze the flow in two stages. First, we describe the dynamics of each gust encounter using persistent homology, a method of topology characterization that identifies the cycles, or ``holes,'' associated with a point cloud~\cite{edelsbrunnerMorozov2014}. Persistent homology has traditionally found success in geometry-focused applications, such as medical imaging~\cite{Qaiser2016} and molecular structure identification~\cite{TownsendMicucciHymelMaroulasVogiatzis2020}. More recent studies have applied persistent homology to the characterization of dynamical events~\cite{MyersMunchKhasawneh2019}, but only a small fraction of these studies consider fluid systems~\cite{kramar2016, green2020, wuTaoZheng2021}. 

Second, we use an autoencoder to transform the full-state dynamics of each gust encounter to a reduced-order space. The autoencoder is a data-driven method of feature extraction, capable of reducing complex fluid flows to a small number of essential state variables~\cite{murataFukamiFukagata2020, fukamiTaira2023}. In its basic configuration, the autoencoder is constructed as an approximation of an invertible, nonlinear transformation between a high-dimensional space and a low-dimensional space. In this work, we construct an autoencoder such that it transforms the dynamics of each gust encounter to a reduced-order space, while also preserving the most prominent topological features of the system. Our goal is to arrive at a low-order space in which the trajectory of the flow is simple and interpretable, all without sacrificing the reconstruction capabilities of the autoencoder. 

In the sections that follow, we present a brief treatise on both persistent homology and nonlinear autoencoders, and how these concepts can be combined to construct topology-preserving maps. We then apply our approach to an experimental gust encounter as a way of demonstrating the utility of this method in compressing, characterizing, and modeling large-scale aerodynamic disturbances.

\section{Methods}
\label{sec:methods}

\subsection{Persistent Homology}

In this section, we review persistent homology and its application to dynamical systems. Persistent homology is a computational tool for identifying homology groups, or $k$-dimensional holes, in a mutlivariate point cloud~\cite{edelsbrunnerMorozov2014}. It characterizes a point cloud based on its topology, or the underlying connectedness of nearby points, and in doing so, provides an avenue for identifying cycles that do not strictly conform to a basis function. 

Mathematically, persistent homology is defined in reference to simplicial complexes. A simplicial complex is a set of interconnected $k$-simplicies, where a 0-simplex corresponds to a vertex, a 1-simplex corresponds to an edge, and a 2-simplex corresponds to a triangle. Under the operation of the symmetric difference, the subcomponents of a simplicial complex, also called $k$-chains, form an algebraic group, and are typically described in terms of their group-theoretic properties. One such property involves the behavior of $k$-cycles, or subcomponents that form $k$-dimensional, undirected closed loops; these $k$-cycles are partitioned into equivalence classes, called homology groups, which express the ``essential'' features of the simplicial complex in a condensed way. The definition of a homology group is $H_k = Z_k / im\big(\partial_{k+1}\big)$, where $Z_k$ denotes the group of $k$-cycles, and $im(\partial_{k+1})$ denotes the image of the boundary operator. In this sense, the $k$-homology group can be interpreted as a basis for the set of all $k$-cycles in a simplicial complex, excluding from that basis the boundaries of $k+1$-simplices, which can be interpreted as noise. The term \textit{persistent} homology refers to the identification of these essential cycles in multivariate point cloud data, where the connectedness of vertices is not known a priori. Persistent homology begins with the \textit{filtration} of point cloud data, or the construction of a sequence of simplicial complexes, and tracks homology groups as they change throughout this sequence~\cite{zomorodian2005}.

\begin{figure}
 \centerline{\includegraphics[width=\textwidth]{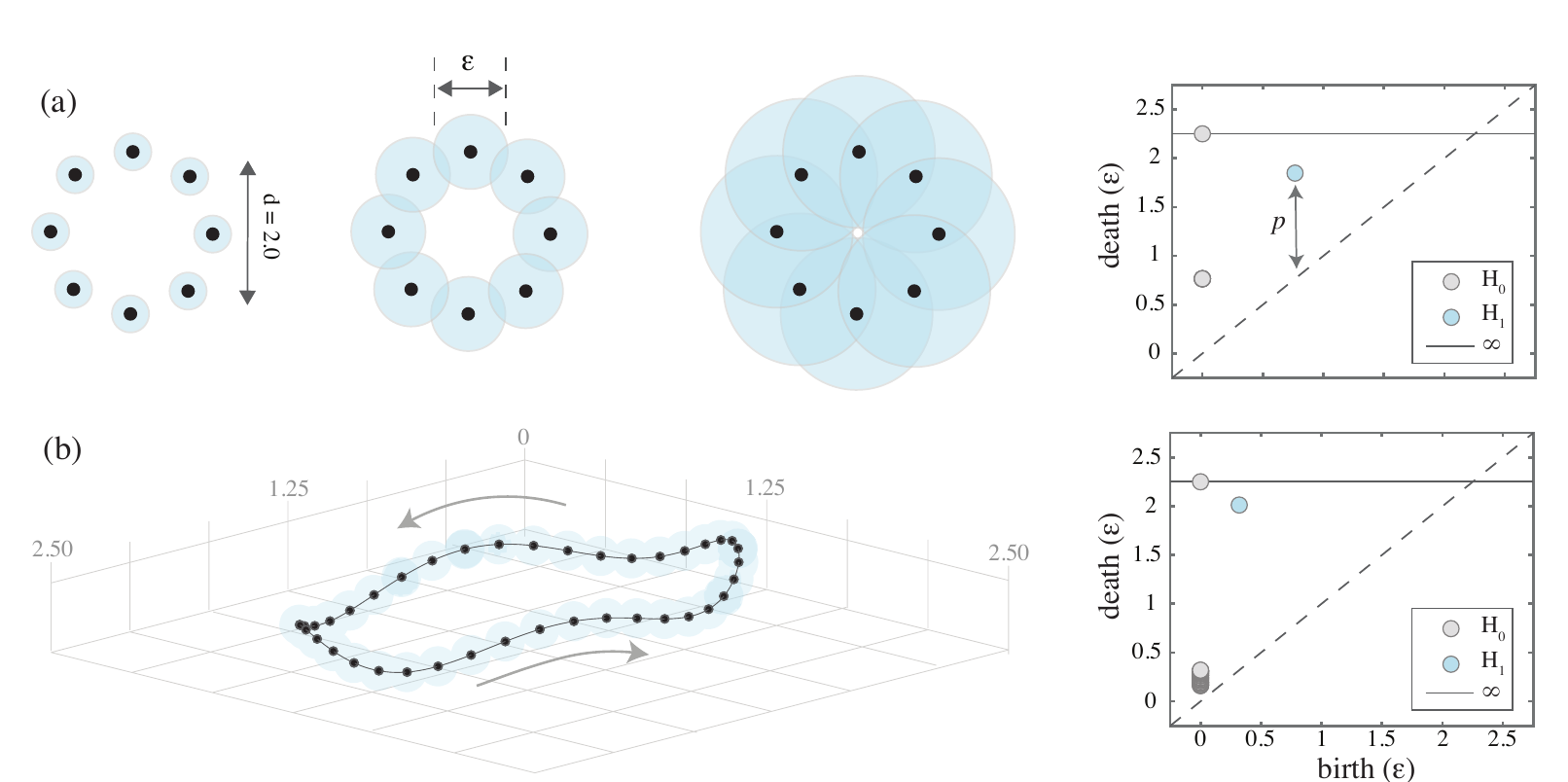}}%
 \caption{The filtration process for a point cloud sampled from a perfect circle (a) and from the trajectory of a representative dynamical system (b).}
 \vspace{-0.65em}
\label{fig2}
\end{figure}

The preceding definition is quite general, but in practice, we can work with a more intuitive understanding of persistent homology. Consider a point cloud sampled from a smooth circle in $\mathbb{R}^2$, sketched in figure~\ref{fig2}(a). Let us associate an $\epsilon$-sphere with each point in figure~\ref{fig2}(a) and examine the intersections that result from gradually increasing its diameter. In particular, we note that there is a certain diameter at which each $\epsilon$-sphere intersects with its neighbors, and a ``hole'' emerges at the center of the point cloud.  Likewise, there is a certain diameter at which all $\epsilon$-spheres share a common intersection, and the hole is closed. The \textit{persistence} of this hole is defined as the difference between the diameter at which the hole emerges (``birth''), and the diameter at which the hole is closed (``death''). 

The procedure above, formally called a Vietoris-Rips filtration, can be used to quantify various topological features, including connected components ($H_0$), holes or 1-cycles ($H_1$), and enclosed voids ($H_2$). These features are typically presented in a persistence diagram, an example of which is provided in figure~\ref{fig2}(a) for the circular point cloud. In this figure, the birth of a topological feature is plotted on the abscissa, and the death of a topological feature is plotted on the ordinate. Collectively, these birth-death pairs provide a concise description of our point cloud in ambient space. The $H_0$ homology group, for instance, is characterized by eight identical birth-death pairs at $(0,1.25)$, each of which corresponds to the intersection of an adjacent $\epsilon$-sphere; meanwhile, the $H_1$ group is characterized by a single birth-death pair at $(0.8,2.30)$, which captures the presence of a single large hole. Note that the dotted line in figure~\ref{fig2}(a) represents minimal persistence (i.e., topological features that emerge and close at the same radius) while the $\mathrm{\infty}$-line represents features that persist to arbitrarily large radii (all persistence diagrams include one $H_0$ element along the $\infty$-line).

As a more practical example, figure~\ref{fig2}(b) plots a series of points sampled from a complex curve in $\mathbb{R}^3$. This curve could be interpreted as the trajectory of a dynamical system, where each axis corresponds to one of three state variables, and each point corresponds to a different instance in time. The rightmost column of figure~\ref{fig2}(b) shows the persistence diagram for this point cloud. We see that much like the circular point cloud, this curve is characterized by one highly persistent hole in $H_1$. The persistence of the hole in figure~\ref{fig2}(b) is quite different from the persistence of the hole in figure~\ref{fig2}(a), but from a top level, the two curves share a similar topological description, in that the rank of their $H_1$ group is identical. This similarity points toward a more powerful idea: the underlying curve in figure~\ref{fig2}(a) is homeomorphic to the unit circle, meaning there exists an invertible map between the two curves.

In the sections that follow, we apply a similar analysis to the dynamics of an unsteady gust encounter. Our core idea is that if we can identify a simple shape that is topologically similar to the full-state dynamics of a gust encounter, then we can construct a map between the two shapes. Because of turbulence and measurement noise, the full-state dynamics of a gust encounter are unlikely to be exactly homeomorphic to such a simple shape~\cite{wuTaoZheng2021}. We aim instead to find a shape that results in a minimal loss of information.

\subsection{Autoencoder}

\begin{figure}
\centerline{\includegraphics[width=\textwidth]{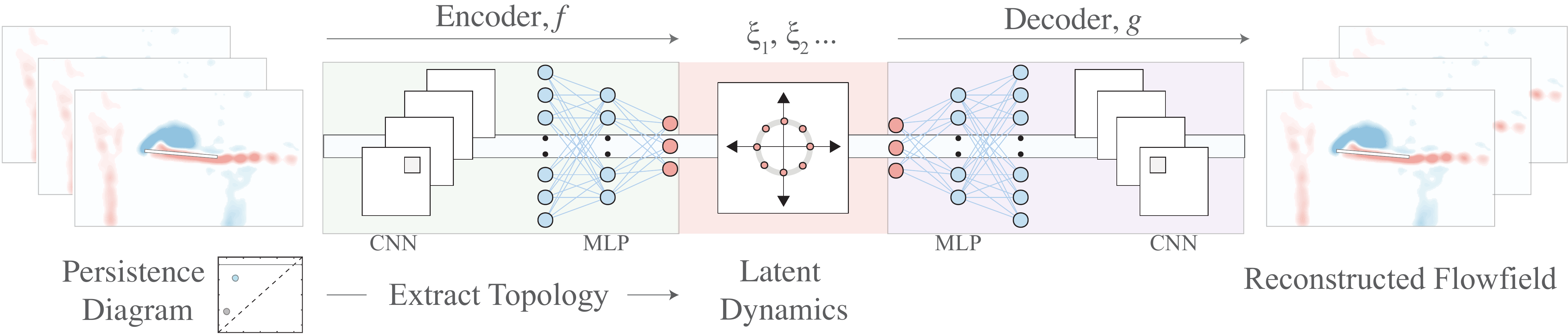}}%
\caption{Schematic of the topological autoencoder.}
\label{fig3}
\end{figure}

In this section, we describe a process for finding transformations between high-dimensional trajectories and simple, low-order shapes, a task we accomplish using an autoencoder. Figure~\ref{fig3} provides a conceptual diagram of our autoencoder. In this figure, a time series of flowfield snapshots is mapped from a high-dimensional state space to a low-order (or latent) space by an encoder $f$ and reverse transformed by a decoder $g$. The encoder and decoder are each composed of a convolutional neural network (CNN, ~\cite{lecun1998}) and a multi-layer perceptron (MLP,~\cite{rumelhart1986}). Within these networks, each layer is associated with a set of model weights, and is coupled to adjacent layers by an activation function, for which we choose the hyperbolic tangent. The latent vector $\mathbf{\xi}$ serves as the bottleneck of the network architecture. The number of components within the latent vector defines the network compression ratio, and typically corresponds to the minimum number of variables from which the full-state can be reconstructed. Stated another way, if the encoder is seen as a nonlinear projection of the full state onto a set of tailored modes, then $\mathbf{\xi}$ represents the temporal coefficients associated with these modes~\cite{murataFukamiFukagata2020}. 

In training an autoencoder, we seek model weights that minimize a loss function, which is typically based on the difference between the input flowfield and the reconstructed flowfield. The key feature of our approach is that in addition to reconstructing the flow, we explicitly instruct the autoencoder to preserve the main topological features of the input trajectory~\cite{moor2020}. Figure~\ref{fig3} provides an illustration of how we incorporate topological constraints into the training of our autoencoder. For a given time series of flowfields, we compute the persistent homology of the trajectory, identify the most persistent topological features, and associate a target shape with the trajectory. Our objective is to select a target shape that retains the main topological features of the input trajectory, while also ensuring that the latent dynamics of the system are mathematically tractable. As an example, the unit circle would serve as a suitable target shape for the curve in figure~\ref{fig2}(b). The two curves are homeomorphic, meaning there exists a lossless transformation between them, and the dynamics of a circular trajectory can be captured by a linear system of ordinary differential equations.

After selecting a shape, we train the autoencoder to minimize the following loss function:

\begin{equation}
\mathcal{L} = \left \| \mathbf{q} - f \circ g (\bf{q}) \right \|_2 + \beta_s \left \| f(\bf{q}) - \mathbf{s}(\kappa) \right \|_2 + \beta_p \left \langle  p \big(f(\bf{q}) \big) - p \big(\mathbf{s}(\kappa) \big) \right \rangle \mathrm{,}
\label{eqn5}
\end{equation}

\noindent where $\mathbf{q}$ represents the full state of the flow; $\mathbf{s}$ represents the trajectory of the target shape (the selection of which depends upon a set of freestream parameters, $\mathbf{\kappa}$); and $p(f(\bf{q}))$ and $p(\bf{s})$ represent the persistence of the latent trajectory and target shape, respectively. Note that vertical brackets in equation~\ref{eqn5} indicate an average among snapshots, while angle brackets indicate an average among trajectories.

Each term in equation~\ref{eqn5} serves a specific purpose in ensuring that the autoencoder converges toward a topology-preserving map. The first term in equation~\ref{eqn5}, called the reconstruction term, ensures that the transformation $f$ is as close as possible to a bijection. Topologically, it ensures that adjacent vertices in $\mathbb{R}^n$ remain adjacent in the latent space. The second term forces the latent trajectory of the system to conform to the target shape ($\mathbf{s}$). Note that this term assumes that $\mathbf{s}$ is given analytically, such that $f(\mathbf{q}) - \mathbf{s}(\kappa)$ is the distance between a point and a curve, but it does not assign a specific location along the shape to each encoded snapshot. The final term in equation~\ref{eqn5} places a minimum threshold on the persistence of holes (or elements of the $H_1$ group) in the latent trajectory, ensuring that $f(\mathbf{q})$ approximates a closed curve for each case. This prevents the autoencoder from converging toward an erroneous local minimum, and can be seen as an aid to convergence.

\section{Results}
\label{sec:results}

We now use our methodology to construct a low-order, data-driven representation of a large aerodynamic disturbance. As a representative dataset, we consider the particle image velocimetry measurements of~\cite{sedkyJones2023}, the basic setup of which is shown in figure~\ref{fig1}. A flat-plate wing (chord $c$ = 7.62 cm, aspect ratio $AR$ = 4,) is towed horizontally at constant velocity ($U_{\infty}$) and constant incidence ($\alpha$) before encountering a transverse, vertical gust. The gust profile is trapezoidal and is characterized by the gust ratio ($G$), or the ratio of gust velocity to wing translational velocity. As the wing is towed through the gust, flow tends to separate about the sharp leading edge of the wing, resulting in the formation and shedding of a leading edge vortex (LEV). In the dataset considered here, the experiment was repeated for three gust ratios ($0.25 \leq G \leq 0.71$) and four incidence angles ($-15^{\circ} \leq \alpha \leq 15^{\circ}$). The variable freestream parameters from equation~\ref{eqn5} are thus defined as $\kappa = \{ \alpha, G\}$.

We begin our analysis by considering a single time series of flowfield measurements corresponding to $\alpha = 5^{\circ}$, $G = 0.71$, and $Re_{\infty} = U_{\infty}c/\nu =10^4$. This case is comprised of 800 snapshots, each consisting of a rectangular, 200 $\times$ 120 grid of spanwise vorticity measurements ($\omega_z$). Note that prior to analysis, we applied a spatial Gaussian filter to each snapshot as a way of normalizing the degree of measurement noise across the time series, which we do not expect to significantly impact our current results.

\begin{figure}
 \centerline{\includegraphics[width=\textwidth]{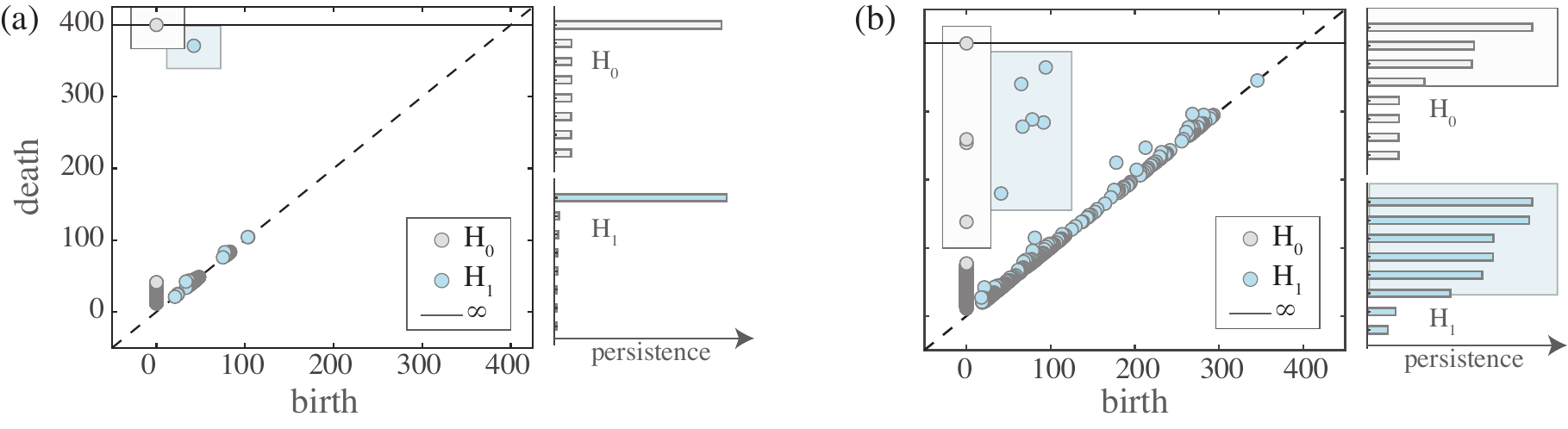}}%
 \caption{Persistence diagram and ${H_0}$/$H_1$ barcodes for (a) the single gust encounter case and (b) the collection of six separate gust encounter cases.}
\label{fig4}
\vspace{-1em}
\end{figure}

Figure~\ref{fig4}(a) shows the persistence diagram associated with this gust encounter. As a first step, let us consider the homology group $H_0$, which tracks connected components throughout the filtration. In figure~\ref{fig4}(a), the $H_0$ group is characterized by a large number of birth-death pairs near $(0,0)$, along with a birth-death pair at the $\infty$-line. These low-persistence pairs can be attributed to the proximity of points associated with adjacent time steps, and are expected for a point cloud sampled from a single continuous trajectory~\cite{MyersMunchKhasawneh2019}. The $H_1$ homology group, which tracks the persistence of holes, provides a much clearer picture of the topology associated with this gust encounter. In figure~\ref{fig4}(a), the $H_1$ group is characterized by a single, highly persistent birth-death pair near $(50, 375)$. The barcodes in figure~\ref{fig4}(a) emphasize the degree to which this single cycle exceeds the persistence of all other $H_1$ group elements. These figures point toward an important observation: because the $H_1$ group contains a single highly persistent element, we expect the data to resembles a single, large loop in state space, with a number of secondary loops emerging along the path. 

The results of figure~\ref{fig4} have important implications in the next step of our analysis. If we were to set a noise threshold in figure~\ref{fig4}, such that we ignore any cycle that lies near the minimum persistence line, then our persistence diagram would consists of a single 1-cycle in $H_1$ and a single connected component in $H_0$. Such a shape is homeomorphic to a simple circle in $\mathbb{R}^2$. We do not suggest that the underlying trajectory is \textit{exactly} homeomorphic to a circle, but knowing that our flow ultimately forms a cycle, we posit that our point cloud can be projected onto this dominant 1-cycle without a significant loss of information. We next assess the validity of this hypothesis, and use the topology-preserving autoencoder to find a transformation between the full-state dynamics of the gust encounter and a simple circle.

\begin{figure}
 \centerline{\includegraphics[width=0.92\textwidth]{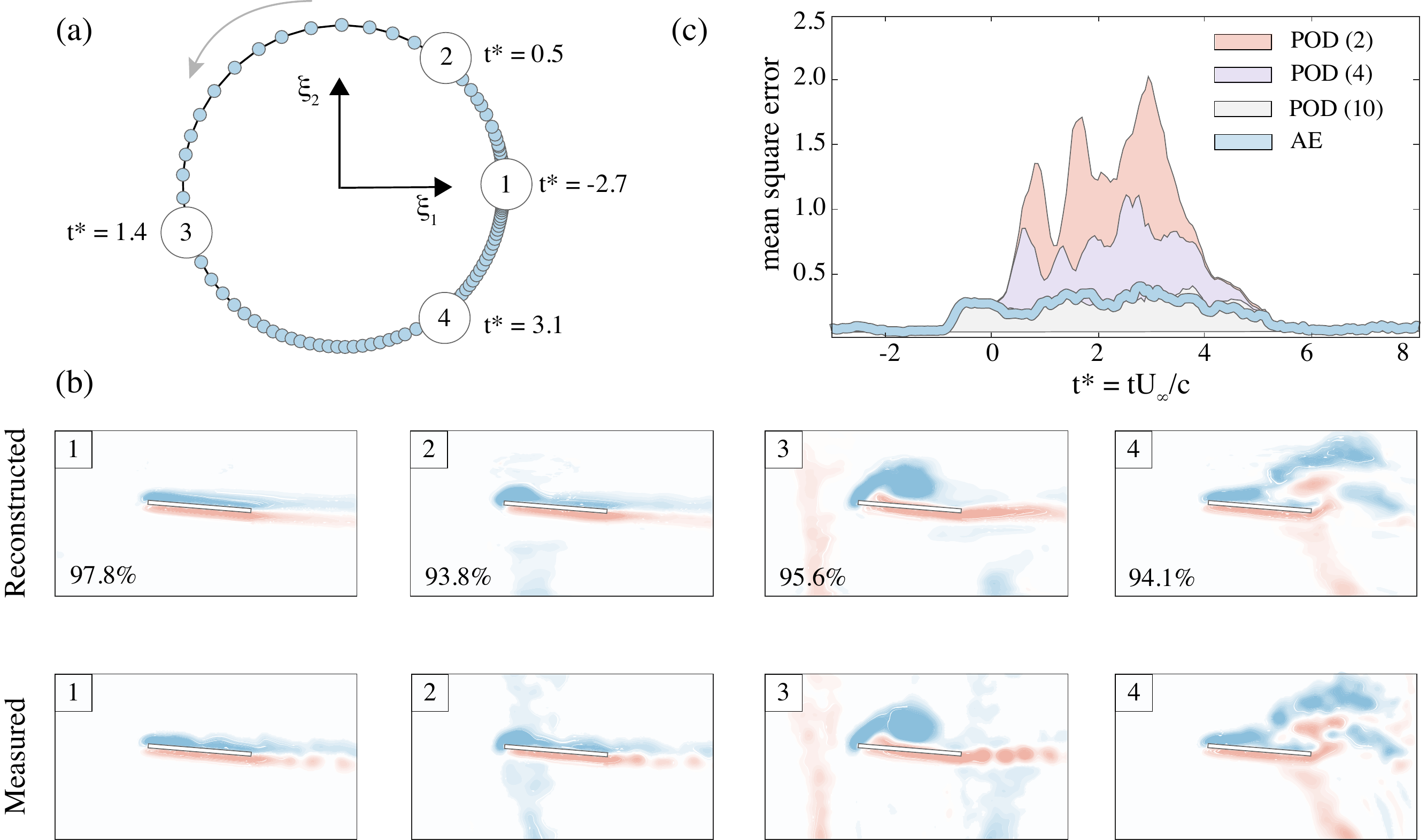}}%
 \caption{(a) The latent space, (b) the reconstructed flowfield, and (c) the mean squared error associated with the $\alpha=5^{\circ}$, $G = 0.71$ case.}
\label{fig5}
\end{figure}

Figure~\ref{fig5} shows (a) the latent space, (b) the reconstructed flowfield, and (c) the relative error that results from encoding the dynamics of the single gust encounter, with 200 snapshots used for training and 600 snapshots used for test/validation. Let us begin by considering figure~\ref{fig5}(a). In this figure, each blue dot corresponds to a state of the system in latent coordinates, and each state is connected to its temporal neighbors by a solid black line. If we start at the base state of the system (state 1), the latent coordinates rotate counter-clockwise as the wing enters the gust (state 2), reach the apogee of the cycle during the shedding of the LEV (state 3), and re-approach the base state during gust recovery (state 4). The evolution of this latent trajectory is roughly continuous with respect to time (i.e., there are no significant jumps between time steps), and thus figure~\ref{fig5}(b) allows us to define a phase angle for each state.

Crucially, it also appears that this cyclic latent space is obtained without a significant loss of flowfield information. Figure~\ref{fig5}(b) compares the reconstructed flowfield, obtained by decoding the latent space, to the original flowfield at key moments throughout the gust encounter. The structural similarity index measure (SSIM, computed with windowing bounds of $-0.05 \leq x/c \leq 2.0$) is included as a quantitative measure of each snapshot's reconstruction accuracy~\cite{wang2004}. Qualitatively and quantitatively, the snapshots in figure~\ref{fig5}(a) reconstruct a range of complex phenomena, including the onset of flow separation, the formation and shedding of an LEV, and the distortion of the gust shear layer upon the wing's exit. The autoencoder struggles somewhat to reconstruct the gust's upstream shear layer, but the near-wing flow features, which ultimately drive the production of aerodynamic force, are accurately reconstructed throughout the entirety of the wing's motion. 

Figure~\ref{fig5}(c) further expounds upon the reconstruction capabilities of our approach, comparing the reconstruction error of our autoencoder with that of proper orthogonal decomposition (POD). In this figure, the topological autoencoder is shown to produce roughly the same error as a ten-mode POD reconstruction, improving upon the compression ratio of POD by a factor of five. Such a favorable comparison indicates that our approach is able to effectively identify the inherent dimension of this gust encounter, and further demonstrates that the simplicity of our latent space does not come at the expense of reconstruction accuracy.

Let us now consider the set of six separate gust encounters. This dataset captures a broader representation of the gust parameter space, and includes a sweep of incidence angles at constant gust ratio ($G = 0.71$, $\alpha = \{ -5^{\circ}, 0^{\circ}, 5^{\circ}, 15^{\circ} \}$) and a sweep of gust ratios at constant incidence ($\alpha = 0^{\circ}$, $G = \{0.25, 0.50, 0.71\}$). As a starting point, figure~\ref{fig4}(b) shows the persistence diagram and $H_0$/$H_1$ barcodes for the set of six gust encounters. This figure plots birth-death coordinates on the abscissa and ordinate, respectively, and was generated by including all six gust encounters in the input point cloud (again, using 200 snapshots from each case for training). Let us first examine homology group $H_0$. In figure~\ref{fig4}(b), the $H_0$ group is characterized by a large number of birth-death pairs near $(0,0)$, and four highly persistent birth-death pairs above $(0,100)$. We can intuit that these four birth-death pairs correspond to our four cases with varying $\alpha$, as a change in $\alpha$ can dramatically affect the state associated with the wing's initial condition. A change in $G$, meanwhile, does not impact the initial condition, and we expect our three $G$ cases to intersect early in the filtration. 

Next, we consider homology group $H_1$. In figure~\ref{fig4}(b), the $H_1$ group is dominated by six highly persistent birth-death pairs, while all remaining birth-death pairs lie near the minimum persistence line. If we classify these low-persistence elements as noise, then we are left with an $H_1$ group of rank 6. As shown in figures~\ref{fig4}(a) and~\ref{fig5}, each individual gust encounter case must be associated with at least one generating cycle, as the wing always re-approaches its base state. The topology of our point cloud can thus be described as a collection of six primary loops, where each loop corresponds to a different combination of $G$ and $\alpha$. 

Let us discuss the implications of figure~\ref{fig4}(b). We have incorporated several new cases into our point cloud, yet the topology described above remains quite simple. In fact, the de-noised topology of figure~\ref{fig4}(b), wherein all states are projected onto the six generating 1-cycles, is homeomorphic to a collection of simple circles in $\mathbb{R}^3$, assuming the relevant intersections among trajectories are maintained. In the final stage of our analysis, we use the topological autoencoder to transform the full-state dynamics of all six gust encounter cases to a series of circles in $\mathbb{R}^3$, the organization of which is manually chosen to be interpretable and tractable. Note that these circular trajectories are embedded in $\mathbb{R}^3$, rather than $\mathbb{R}^2$, as a way of avoiding non-physical intersections between latent trajectories.

\begin{figure}
\centerline{\includegraphics[width=\textwidth]{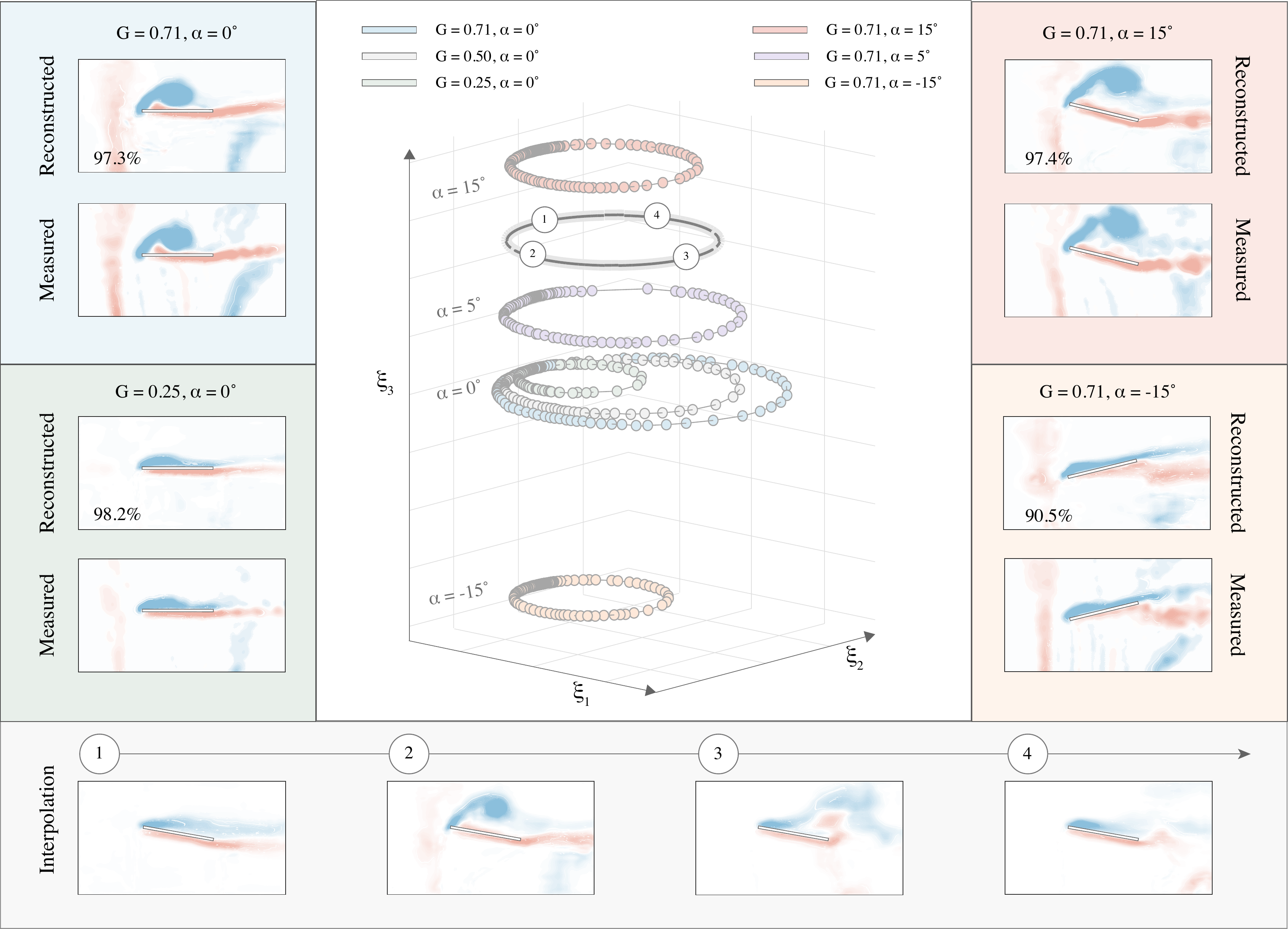}}%
\caption{The latent space for the set of six gust cases. Insets show snapshots of the reconstructed flowfield, each with a corresponding structural similarity index measure.}
\label{fig7}
\end{figure}

Figure~\ref{fig7} shows the latent space identified by the topological autoencoder. Each color-coded trajectory corresponds to a different combination of $G$ and $\alpha$, and each marker corresponds to a different state of the flow. In constructing this figure, we specifically constrained the latent variables ($\xi_1, \xi_2, \xi_3$) such that (1) a change in vertical location can be interpreted as a change in incidence, (2) a change in circle radius can be interpreted as a change in the level of unsteadiness, and (3) the latent dynamics of each gust encounter evolve on a $\xi_1$-$\xi_2$ hyperplane. With these constraints, we arrive at a latent space in which each circle is individually interpretable from a cyclic perspective (i.e., its phase is continuous with respect to time), and in which the space as a whole can be readily linked to physical events. Note that we encoded all six gust encounters simultaneously, using 200 snapshots from each case as a training dataset. Likewise, we chose the radius of each circle such that its persistence in figure~\ref{fig7} is proportional to its persistence in the full-state.

On its own, figure~\ref{fig7} is an intuitive way of visualizing multiple gust encounter cases from the perspective of loops and cycles, but the notable feature of figure~\ref{fig7} is that this latent space can be decoded to accurately recreate the original flowfield. The insets of figure~\ref{fig7} show a series of comparisons, both qualitative and quantitative, between the reconstructed vorticity field and the measured vorticity field for four representative gust encounter cases. Each pair of flowfields represents a different combination of $G$ and $\alpha$ and corresponds to a phase at which the wing is completely immersed in the gust. These cases cover a wide spectrum of physical phenomena; the case with negative incidence ($\alpha = -15^{\circ}$, $G = 0.71$), for instance, exhibits a reversal in the orientation of its leading shear layer prior to the growth of an LEV. Figure~\ref{fig7} indicates that the autoencoder is able to accurately reconstruct the vortex shedding behavior of each of these various cases by simply decoding a collection of circular trajectories in $\mathbb{R}^3$.

While the coverage of the subspace shown in figure~\ref{fig7} is limited by data availability, this figure still has fairly far-reaching implications, from analysis to reduced-order modeling to state estimation. As a demonstration, the bottom row of figure~\ref{fig7} shows an interpolated time series of flowfield snapshots extracted from our simple latent space. These snapshots were generated by drawing a new circular trajectory in the latent space, with an $\xi_3$ coordinate that lies between two of our training cases ($\alpha = 5^{\circ}$ and $15^{\circ}$). In these interpolated snapshots, we see that the wing exhibits an incidence angle of $\alpha \approx 10^{\circ}$, and the vorticity field evolves in a physically realizable manner, capturing the expected stages of LEV growth and shedding. 

\begin{figure}
\centerline{\includegraphics[width=\textwidth]{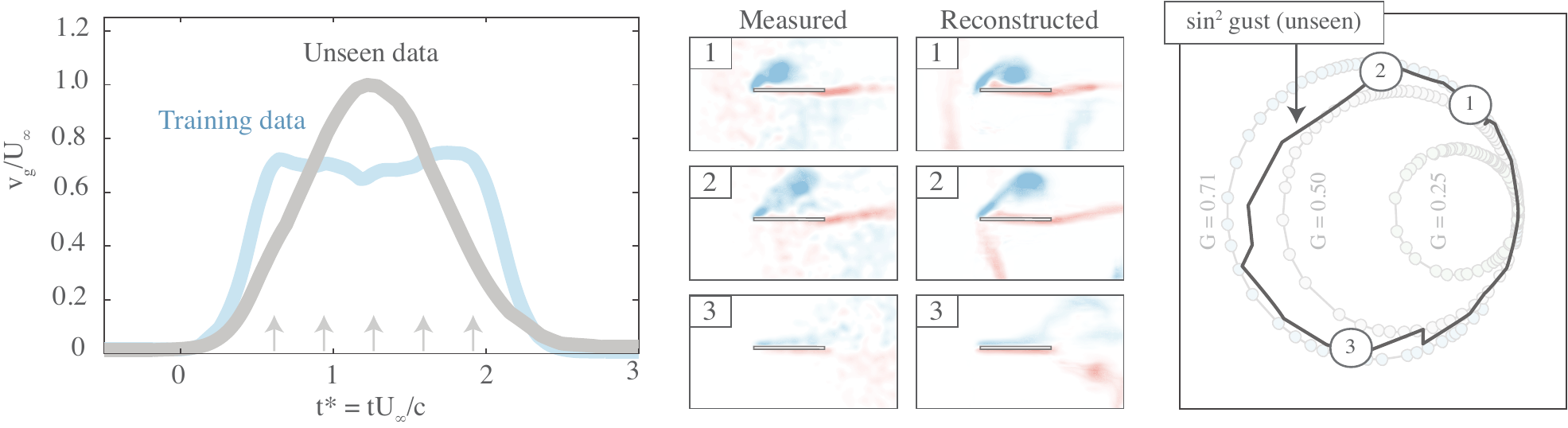}}%
\caption{Application of the autoencoder to an unseen gust encounter case, showing (left) the gust profile, (center) the flowfield reconstruction, and (right) the latent trajectory.}
\label{fig8}
\vspace{-1em}
\end{figure}

Figure~\ref{fig8} further explores the link between our latent space and the physical evolution of the flowfield. In this figure, we apply the autoencoder to a gust encounter case that was not included in the training dataset, as a way of demonstrating physical relevance and utility of our latent coordinates. This untrained case consists of a transverse gust encounter at $\alpha=0^{\circ}$ and $G = 1.0$; it was measured in the same water tank facility as our training data but features a different gust profile and a larger gust ratio~\cite{towne2023}. Figure~\ref{fig8} shows that our autoencoder is able to reconstruct the location and size of the primary vortex in the untrained case with a very reasonable degree of accuracy. In addition, the latent trajectory associated with this case (right side of figure~\ref{fig8}) indicates that the $\mathrm{sin^2}$ gust at $G = 1.0$ is actually quite similar to a trapezoidal gust at $0.5 < G < 0.71$, suggesting that the mean gust velocity (which is similar between the two profiles) may represent a condensed metric for predicting the evolution of the LEV. We can thus reasonably conclude that the latent space of figure~\ref{fig7} represents more than a way of compressing an unsteady flow. Rather, it demonstrates that our cyclic approach, wherein the dynamics of the flow are transformed to a simple (yet topology-preserving) trajectory, is capable of identifying subspaces in which complex flows can be easily modeled, interpolated, and understood.

\section{Concluding Remarks}

We considered a cyclic approach to the study of discrete gust encounters, focusing on experiments in which a wing is towed through a transverse gust. We used persistent homology to describe each gust encounter as a cyclic event, and determined that each case could be described by a single, highly persistent 1-cycle. We then posited that because of their topological simplicity, each gust encounter could be transformed to a basic shape. Using a nonlinear autoencoder, we identified a subspace in which the dynamics of six separate gust cases are represented by six circular trajectories in $\mathbb{R}^3$; these circular trajectories can be decoded to accurately reconstruct the original flowfield. The current approach holds promise in its ability to identify transformations suited to low-order modeling, and our future work aims to generalize these transformations for a variety of flow configurations and applications.

\section*{Acknowledgements}
The authors wish to thank A. Linot for many helpful discussions on nonlinear dynamics and topology. {\bf Funding:} L.S., K.F., and K.T. thank the generous support of the US Department of Defense Vannevar Bush Faculty Fellowship (Grant No. N00014-22-1-2798) and the US Air Force Office of Scientific Research (Grant No. FA9550-21-1-0178). G.S. and A.J. thank the generous support of the US Air Force Office of Scientific Research (Grant No. FA9550-16-1-0508) and the National Science Foundation (Award No. 2003951). K.F. acknowledges support from the UCLA-Amazon Science Hub for Humanity and Artificial Intelligence. {\bf Declaration of interests:} The authors report no conflict of interest.

\bibliographystyle{unsrt}
\bibliography{manuscript}

\end{document}